\documentclass[aps,prl,twocolumn,amsmath,amssymb,superscriptaddress,preprintnumbers]{revtex4}
\usepackage{graphicx}
\usepackage{amssymb}
\usepackage{natbib}
\usepackage{xfrac}
\usepackage{epstopdf}
\usepackage{float}
\usepackage{color}

\begin{document}

\title{Plaquette instability competing with bicollinear ground state in detwinned FeTe}

\author{David W. Tam}
\affiliation{Department of Physics and Astronomy, Rice University, Houston, Texas 77005, USA}
\author{Hsin-Hua Lai}
\affiliation{Department of Physics and Astronomy, Rice University, Houston, Texas 77005, USA}
\author{Jin Hu}
\affiliation{Department of Physics, University of Arkansas, Fayetteville, Arkansas 72701, USA.}
\author{Xingye Lu}
\affiliation{Department of Physics, Beijing Normal University, Beijing 100875, China}
\author{H. C. Walker}
\affiliation{ISIS Facility, Rutherford Appleton Laboratory, STFC, Chilton, Didcot OX11 0QX, United Kingdom}
\author{D. L. Abernathy}
\affiliation{Quantum Condensed Matter Division, Oak Ridge National Laboratory, Oak Ridge, Tennessee 37831, USA}
\author{J. L. Niedziela}
\affiliation{Neutron Scattering Division, Oak Ridge National Laboratory, Oak Ridge, Tennessee 37831, USA}
\author{Tobias Weber}
\affiliation{Institut Laue-Langevin, 71 avenue des Martyrs, CS 20156, 38042 Grenoble cedex 9, France}
\author{M. Enderle}
\affiliation{Institut Laue-Langevin, 71 avenue des Martyrs, CS 20156, 38042 Grenoble cedex 9, France}
\author{Yixi Su}
\affiliation{J{\"u}lich Centre for Neutron Science, Forschungszentrum J{\"u}lich GmbH, Outstation at MLZ, D-85747 Garching, Germany}
\author{Z. Q. Mao}
\affiliation{Department of Physics, The Pennsylvania State University, University Park, Pennsylvania 16802, USA}
\author{Qimiao Si}
\affiliation{Department of Physics and Astronomy, Rice University, Houston, Texas 77005, USA}
\author{Pengcheng Dai}
\email{pdai@rice.edu}
\affiliation{Department of Physics and Astronomy, Rice University, Houston, Texas 77005, USA}
\affiliation{Department of Physics, Beijing Normal University, Beijing 100875, China}

\date{\today}

\begin{abstract}
We use inelastic neutron scattering to show that long-range spin waves arising from the static bicollinear antiferromagnetic (AF) order in FeTe, which have twofold rotational symmetry in a fully detwinned crystal, 
rapidly dissolve above  $E\approx 26$ meV into ridges of scattering with fourfold rotational symmetry 
and a nearly isotropic magnetic fluctuation spectrum.
With increasing temperature above $T_N\approx 68$ K, the twofold spin waves change into 
broad regions of scattering with fourfold symmetry. 
Since the scattering patterns from plaquette magnetic order generated within a bilinear biquadratic Hamiltonian
have fourfold rotational symmetry consistent with the high-energy, spin-isotropic spin waves of FeTe, 
we conclude that the bicollinear AF 
state in FeTe is quasidegenerate with plaquette magnetic order, 
providing evidence for the strongly frustrated 
nature of the local moments in iron chalcogenide family of iron-based superconductors. 
\end{abstract}

\maketitle

\section{Introduction}

Superconductivity in iron-based superconductors arises from long-range ordered antiferromagnetic (AF) parent compounds \cite{scalapinormp,stewart,dairmp,si_natrevmat}.
Magnetism is thus believed to play a central role in the mechanism of high-temperature (high-$T_c$) 
superconductivity.  Correspondingly,
 it is crucial to determine the magnetic ground state 
of the parent compounds of iron-based superconductors.
For iron pnictides such as $A$Fe$_2$As$_2$ ($A=$ Ba, Ca, Sr), a parent of iron pnictide superconductors, the material exhibits a tetragonal-to-orthorhombic structural transition at $T_s$ and forms twin-domains before ordering antiferromagnetically at $T_N$ ($T_s\ge T_N$) \cite{qhuang,mgkim}.
Although the collinear AF structure of the parent compounds exhibits twofold ($C_2$) rotational symmetry \cite{qhuang,mgkim}, the formation of twinned domains in the AF ordered state means that experimental measurements of spin waves have fourfold ($C_4$) rotational symmetry and therefore do not allow a conclusive determination of the magnetic ground state \cite{zhaojun,ames,leland11,Ewings11}.
By carrying out inelastic neutron scattering experiments on 100\% mechanically detwinned samples \cite{Dhital12,Lu14}, spin waves in BaFe$_2$As$_2$ were found to change from $C_2$ to $C_4$ symmetry starting at energies near 150 meV and extending to the top of the magnetic bandwidth, establishing the magnetic ground state of the system \cite{xylu18}.

FeTe, which is a parent of iron chalcogenide superconductors, exhibits coupled bicollinear (BC) AF order and a tetragonal-to-monoclinic structural transition below $T_s=T_N\approx 68$ K and a low temperature ordered moment of $\sim$2.0 $\mu_B/$Fe [Fig. 1(a)] \cite{bao09,sli09,Rodriguez11}. Its magnetic ground state has been described as either a local moment Heisenberg Hamiltonian, which has $C_2$ symmetry \cite{LCFP11}, or as spin-liquid polymorphism, where a short-range ferromagnetic (FM) plaquette phase with $C_4$ symmetry coexists with a short-range AF plaquette (PL) phase with $C_2$ symmetry \cite{ZXWT12,ZXTG11,ZSLT15}.
In a bilinear biquadratic model containing coupling terms up to the third-nearest-neighbor, it has been argued that the bicollinear order is degenerate with plaquette order \cite{Lai16}, where in the low-temperature AF ordered state, the bicollinear order prevails over the plaquette order due to the presence of ring exchange \cite{Lai16} or spin-lattice coupling \cite{Bishop16}.
Since previous neutron scattering experiments in the AF ordered state were carried out on twinned samples with $C_4$ rotational symmetry \cite{LCFP11,ZXWT12,ZXTG11,ZSLT15,Parshall,Stock14,stock17}, it is unclear if the intrinsic spin excitations in detwinned FeTe have $C_2$ or $C_4$ symmetry and if any of these models contain the key ingredients for a successful understanding of the physics \cite{Lai16,Bishop16}.

\begin{figure}[t]
\includegraphics[scale=.42]{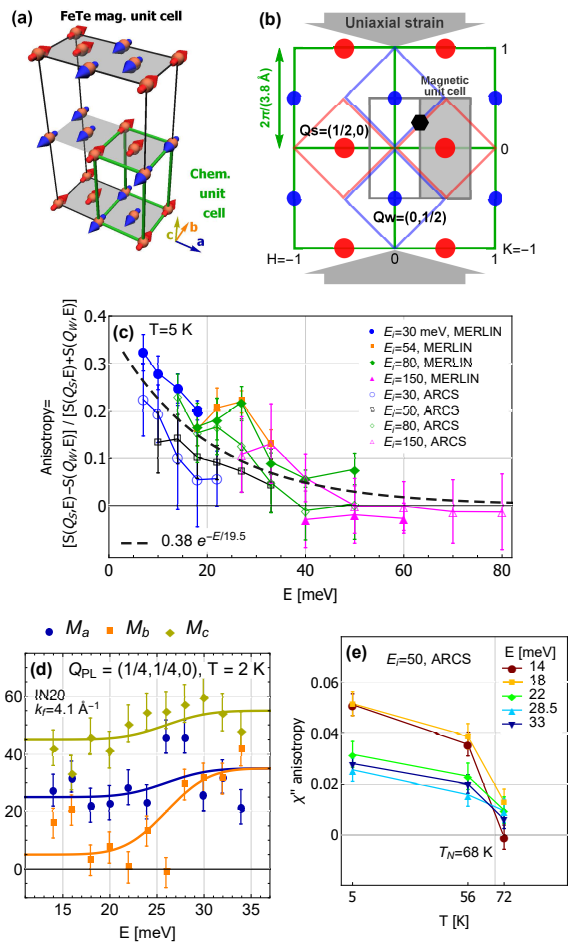}
\caption{
(a) Real-space spin structure of AF ordered FeTe (only Fe atoms are shown).
Color arrows indicate lattice directions and directions of spin excitations \cite{Ysong18}. 
(b) ${\bf Q}$-space configuration, showing the magnetic unit cell in grey, the majority-domain Bragg reflections $\mathbf{Q}_{s}=(\pm \sfrac{1}{2},0)$ in red dots for pressure applied along the $b$-axis, and the minority-domain Bragg reflections $\mathbf{Q}_{w}=(0,\pm \sfrac{1}{2})$ as smaller blue dots. The diamond-shaped zones integrated for calculating the anisotropy ratio in part (c) are shown in red/blue.
(c) Anisotropy of the integrated intensity in the four diamond-shaped zones at $T=5$ K, from time-of-flight neutron scattering, demonstrating a complete relaxation to fourfold symmetry by $E=40$ meV. (d) Magnetic fluctuation spectrum for the plaquette position ${\bf Q}_{\rm PL}=(\sfrac{1}{4},\sfrac{1}{4},0)$ shown as a black hexagon in (b). Fluctuations become nearly isotropic above $E\approx 26$ meV.
(e) Temperature dependence of the anisotropy between strong and weak zones, as a function of energy. The anisotropy drops off rapidly near 22 meV.
}
\label{fig1}
\end{figure}

\begin{figure*}[t]
\includegraphics[scale=.42]{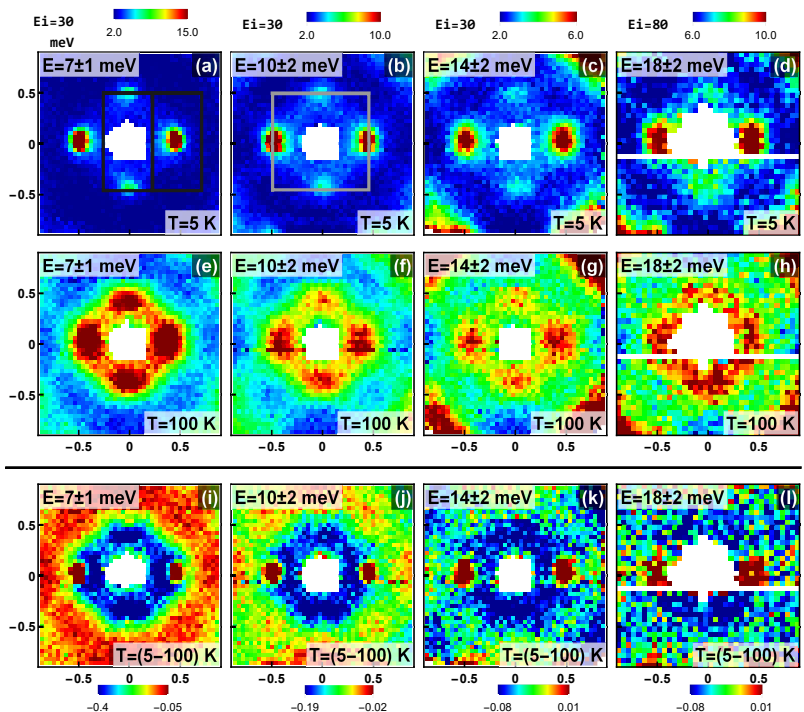}
\caption{ 
Neutron scattering data from MERLIN at constant energy transfer, at $T=5$ K (a-d) and 100 K (e-h). The color scale corresponding to both temperatures is shown above the column. (i-l) Slices from the direct subtraction of the data sets at 5 K $-$ 100 K with color scales shown below, with red regions corresponding to increased intensity in the low-temperature phase at (0,\sfrac{1}{2}) and crystallographically equivalent positions. The black boxes in (a) show two magnetic Brillouin zones [Fig. \ref{fig1}(b)], and the grey box in (b) shows the central region plotted in Fig. \ref{fig:exp-th}.
}
\label{fig:tof-lowE}
\end{figure*}

\begin{figure}[t]
\includegraphics[scale=.42]{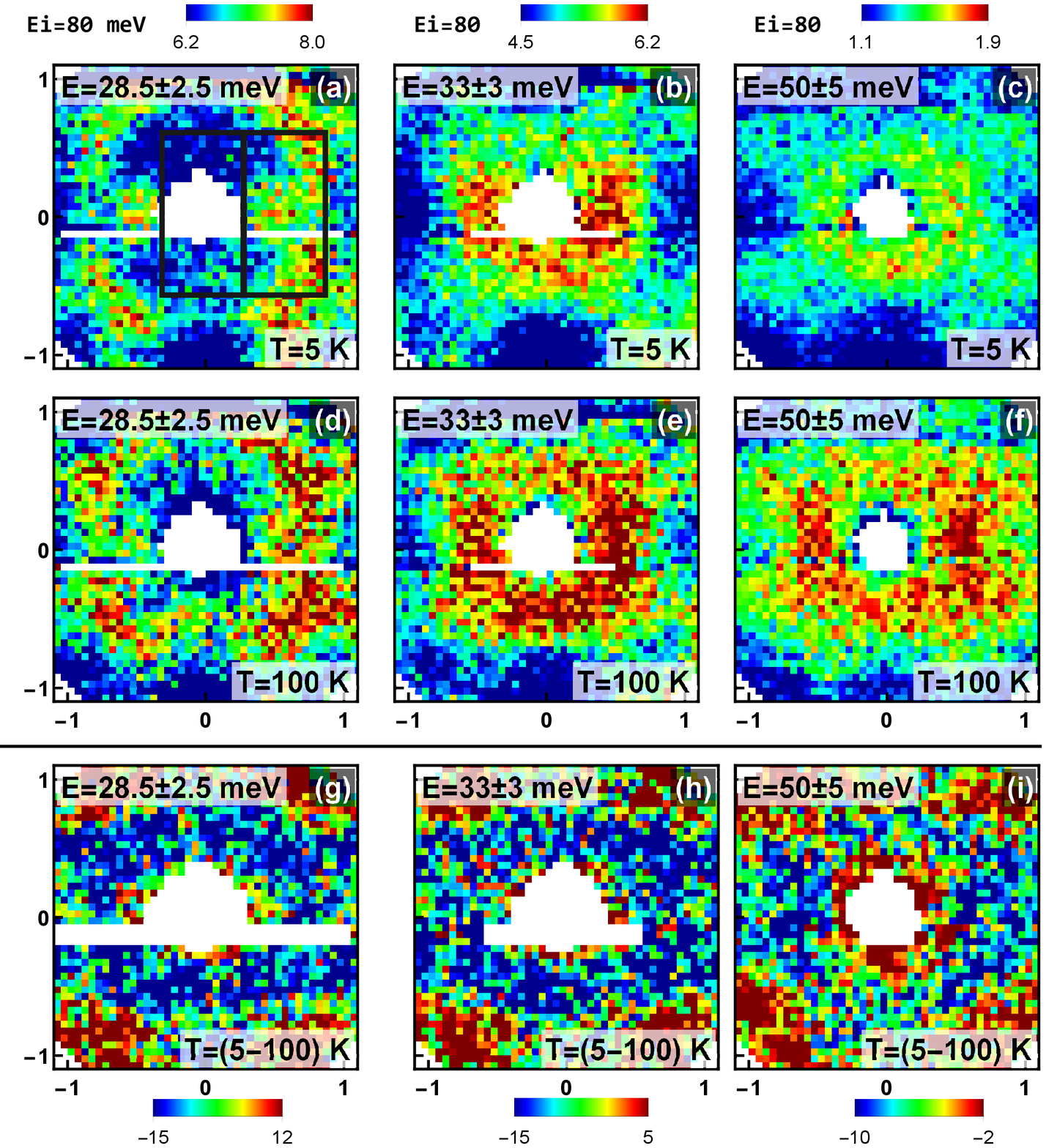}
\caption{
Neutron scattering data from MERLIN at constant energy transfer, at $T=5$ K (a-c) and $T=100$ K (d-f) with the corresponding color scale for both temperatures shown above the column. (g-i) Slices from the direct subtraction of the datasets at 5 K $-$ 100 K with color scales shown below. The white areas represent the detector gaps and the direct beam block.
}
\label{fig:tof-highE}
\end{figure}

In this article, we report inelastic neutron scattering studies of spin excitations in uniaxial pressure detwinned FeTe with $T_s=T_N\approx 68$ K [Fig. 1(b)] \cite{Ysong18}.
At low energies, we find a long-range bicollinear AF phase, exhibiting spin waves with $C_2$ symmetry at AF wave vectors ${\bf Q}_{\rm BC}=(m\pm 1/2,n)$ for integer $m,n$ (Fig. \ref{fig:tof-lowE}). These $C_2$-symmetric spin waves coexist with weak $C_4$-symmetric spin excitations consistent with a model of FM plaquette order where excitations are broad ridges of scattering around ${\bf Q}_{\rm PL}=(m,n)$ \cite{ZXWT12,ZXTG11,ZSLT15}.
Surprisingly, as a function of increasing energy, the $C_2$ anisotropy rapidly decays with increasing energy [Fig. 1(c)], indicating true frustration in the magnetic ground state of FeTe between twofold and fourfold rotational symmetric ground states.
Above $E\approx 26$ meV, the entire excitation spectrum adopts $C_4$ symmetry [Figs. \ref{fig1}(c) and \ref{fig:tof-highE}].
Our neutron polarization analysis confirms the magnetic nature of the $C_2$ and $C_4$ symmetric excitations, and reveals that both excitations have similar spin space anisotropy [Fig. \ref{fig1}(d)].
Conducting identical measurements in the high temperature tetragonal phase ($T=100$ K), we find the bicollinear phase is suppressed, with spin excitations appearing as paramagnetic isotropic scattering with $C_4$ rotational symmetry down to 8 meV [Figs. \ref{fig1}(e), \ref{fig:tof-lowE} and \ref{fig:tof-highE}].
We understand these results by modeling our data with a Hamiltonian consisting of competing bicollinear and plaquette instabilities \cite{Lai16}, and conclude that the Fe local moments are strongly frustrated in the iron chalcogenide family of iron-based superconductors \cite{WSPZ16,RYu15,Fwang15,Glasbrenner}.
Previously, it was observed that interstitial Fe in superconducting 
Fe$_{1+y}$Te$_{1-x}$Se$_x$ can induce a magnetic Friedel-like oscillation 
that arises from the interstitial
moment surrounded by compensating FM plaquette order, and may seed the double-stripe order found in FeTe \cite{Thampy12}.
The success of our model suggests that the frustration is intrinsic to the FeTe layers, and may not be driven by the interstitial iron moments.

\section{Experimental Results}

\subsection{Neutron Time-Of-Flight Spectroscopy}

Our experiments were carried out at ARCS and MERLIN neutron time-of-flight chopper spectrometers at the Spallation Neutron Source, Oak Ridge National Laboratory, and ISIS, Rutherford-Appleton Laboratory, respectively.
We define the wave vector \textbf{Q} in three-dimensional reciprocal space in \AA$^{-1}$ as ${\bf Q}=H {\bf a^\ast}+K{\bf b^\ast}+L{\bf c^\ast}$, where $H$, $K$, and $L$ are Miller indices and ${\bf a^\ast}=\hat{{\bf a}}2\pi/a, {\bf b^\ast}=\hat{{\bf b}} 2\pi/b, {\bf c^\ast}=\hat{{\bf c}}2\pi/c$ are reciprocal lattice units (r.l.u.) of the crystallographic unit cell [Fig. \ref{fig1}(b)].
In the low-temperature AF ordered phase of FeTe, $a\approx 3.8$ \AA, $b\approx 3.8$ \AA, and $c\approx 6.24$ {\AA} \cite{Ysong18}. 
Measurements were carried out with incident neutron energies $E_i=30, 50, 54, 80$, and 150 meV with the incident beam fixed along the $c$ axis.
Using this fixed geometry, $L$ values are completely determined by the choice of neutron transfer energy at a given neutron incident energy $E_i$; and one cannot make arbitrary choice of both energy and $L$.
For this reason, we have sometimes abbreviated the wavevector $\mathbf{Q}=(H,K,L)$ in the text to simply $(H,K)$ when the energy and $E_i$ are also specified.
Consistent with earlier work \cite{LCFP11}, we find that there is little $L$-dependence in this layered material for energy transfers above two or three times the spin gap energy.
Data analysis was conducted by reducing the event data sets into four-dimensional coordinate space using the neutron analysis software Mantid, then projected into constant-energy slices using Mslice in Matlab.

We use an aluminum mount with stainless steel spring washers, as described in previous publications \cite{xylu18}, to achieve 
75\% detwinning of FeTe single crystals (7:1 domain population). Aluminum plates fit inside tracks in the manifold and are compressed against the FeTe crystals by tightening aluminum screws, and the compression is held by a stack of spring washers. In the MERLIN experiment, the spring washers and a large portion of the aluminum outside the central area was blocked with two pieces of neutron-absorbing B$_4$C that were fastened to the manifold with aluminum wire. In the ARCS experiment, we blocked the spring washers with Cd foil.

\begin{figure}[t]
\includegraphics[scale=.45]{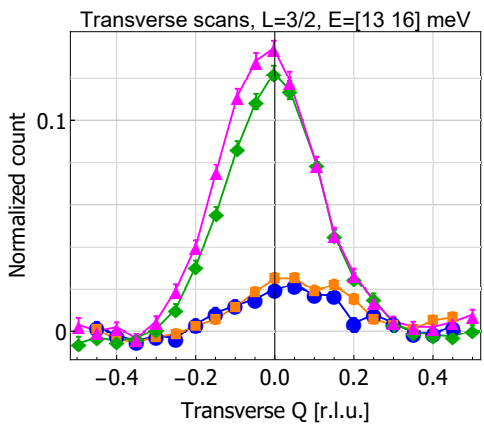}
\caption{
Transverse cuts along the magnetic positions, at $E=14.5\pm 1.5$ meV, $L=1.5$, used to calculate the detwinning ratio at MERLIN.
}
\label{fig:tvscans}
\end{figure}

To determine the detwinning efficacy in this clamp device, we first project data from the $E_i=30$ meV dataset into a constant-energy slice containing energies $13<E<16$ meV. This data was carefully chosen to satisfy the condition $L=1.5$, which is a magnetic Bragg peak position \cite{bao09,sli09}. To calculate the detwinning ratio, we then integrate in the crystallographic longitudinal direction over $0.3<H<0.6$ for transverse cuts along the $K$ direction, and likewise integrate over $0.3<K<0.6$ for transverse cuts along the $H$ direction. We then project these transverse cuts along one-dimensional (1D) axes and subtract the background from the peak positions by choosing points far away from the central region containing the peak intensity. The resulting background-subtracted 1D cuts are shown in Fig. \ref{fig:tvscans}. Finally, we combine the intensity from the two strong ($s$) positions and the two weak ($w$) positions and integrate the remaining intensity, finding $(I_s-I_w)/(I_s+I_w) = 0.75$.
We note that this detwinning ratio is the lower bound of the actual detwinning ratio, based on the 
assumption that there should be no spin excitations near the weak Bragg peak position for 100\% detwinned FeTe. 

From the applied uniaxial pressure direction, shown in Fig. \ref{fig1}(b), we expect that static AF order in detwinned FeTe occurs in a double stripe configuration along the second-nearest neighbor direction, which is the $(0,\pm\sfrac{1}{2})$ position within the two-iron tetragonal chemical unit cell.
This position differs from the iron pnictides, which form single stripe AF order at the $(\pm \sfrac{1}{2},\pm \sfrac{1}{2})$ positions in this notation, and from FeSe, which exhibits no magnetic order but demonstrates low-energy fluctuations near $(\sfrac{1}{2},\sfrac{1}{2})$ and $(\sfrac{1}{2},-\sfrac{1}{2})$ in a twinned sample \cite{WSPZ16}.

To demonstrate how the low-energy spin excitations evolve with increasing energy in the AF ordered state, we calculate the spin excitation anisotropy $\eta=(I_{s}-I_{w})/(I_{s}+I_{w})$, where $I_s=I(\pm \sfrac{1}{2},0)$ and $I_w=I(0,\pm \sfrac{1}{2})$ are the integrated intensity inside the two red and two blue diamond-shaped zones located symmetrically around (0,0) as shown in Fig. \ref{fig1}(b).
Each diamond has the area of one magnetic unit cell, so the total area encompassing both strong peaks or both weak peaks is two magnetic unit cells.
We find that the spin excitation anisotropy decays rapidly with increasing energy, and as a guide to the eye we make an exponential fit [Fig. \ref{fig1}(c)] with an energy scale of about 20 meV.

Figure \ref{fig:tof-lowE} summarizes two-dimensional (2D) images of spin excitations of detwinned FeTe as a function of increasing energy in the AF ordered phase at $T=5$ K and at the paramagnetic tetragonal phase at $T=100$ K.
At $E=7\pm 1$ meV [Fig. \ref{fig:tof-lowE}(a)], we see clear anisotropic spin excitations with strong peaks at the expected bicollinear AF ordering position $\mathbf{Q}_{s}=(\pm \sfrac{1}{2},0)$ and smaller peaks at $\mathbf{Q}_{w}=(0,\pm \sfrac{1}{2})$.
In addition, we see ridge-like scattering connecting the AF positions in a diagonal fashion.
Upon increasing energies of spin excitations to $E=10\pm 2$, $E=14\pm 2$, 
and $18\pm 2$ meV, the anisotropy between strong and weak positions becomes progressively smaller, and the ridge scattering increases [Figs. \ref{fig:tof-lowE}(b)-\ref{fig:tof-lowE}(d)].
On warming to $T=100$ K well above $T_N$ and $T_s$, the anisotropic scattering in the AF ordered state in Figs. \ref{fig:tof-lowE}(a)-\ref{fig:tof-lowE}(d) exhibits a predominantly $C_4$ symmetric pattern as shown in Figs. \ref{fig:tof-lowE}(e)-\ref{fig:tof-lowE}(h).
Figures \ref{fig:tof-lowE}(i)-\ref{fig:tof-lowE}(l) show temperature-subtracted data between 5 K and 100 K, which strikingly reveals the difference between the incommensurate instability at $\mathbf{Q} \approx (\pm 0.42,0)$ and$(0,\pm 0.42)$ in the high-temperature paramagnetic phase and the commensurate spin waves in the low temperature phase \cite{Parshall}.

Similarly, Figures \ref{fig:tof-highE}(a)-\ref{fig:tof-highE}(c) show 2D images of spin excitations in the AF ordered state at $E=28.5\pm 2.5$, $33\pm 3$, 
and $50\pm 5$ meV, respectively.
At these higher energies, we see that the $C_2$ symmetry of the low-energy spin excitations is essentially replaced by $C_4$ symmetry.
For $E=30$ meV and above, spin excitations form ridge-like scattering near the origin with four lobes extending to $(\pm \sfrac{1}{2},\pm \sfrac{1}{2})$ [Figs. \ref{fig:tof-highE}(b) and \ref{fig:tof-highE}(c)].
These ridge-like excitations persist to the top of the magnetic bandwidth near $E\sim$250 meV \cite{LCFP11} and give rise to the intensity minima observed at the $(1,0)$ and $(1,1)$ positions in all the constant-energy slices shown in Fig. \ref{fig:tof-highE}.
On warming to $T=100$ K well above $T_N$ and $T_s$, the overall scattering intensity increases but the wavevector dependence remains essentially unchanged [Figs. \ref{fig:tof-highE}(d)-\ref{fig:tof-highE}(f)].
Figures \ref{fig:tof-highE}(g)-\ref{fig:tof-highE}(i) show temperature differences between 5 K and 100 K, which reveals some intensity increase at low temperature along directions 45 degrees from the uniaxial pressure direction, but has no signature of the 2D 
spin waves seen at lower energies.

Overall, the neutron scattering data shows that spin excitations in detwinned FeTe evolve between two behaviors.
Below $T_N$, spin excitations below 20 meV exhibit sharp peaks at $(\pm\sfrac{1}{2},0)$ that only slightly broaden into cones with increasing energy, consistent with transverse spin wave nature of the excitations associated with AF bicollinear order \cite{Ysong18}. 
For energies above 30 meV, additional spin excitations appear that consist of a ridge-like scattering reminiscent of the patterns observed in FeTe$_{1-x}$Se$_{x}$ \cite{LCGN10}.
The most important features to be captured by a theoretical model of FeTe are therefore clear: the model must demonstrate twofold symmetry in its ground state, and rapidly transition to fourfold symmetry at very low energies compared to the magnetic band top of nearly 250 meV \cite{LCFP11}.

\subsection{Polarized Neutron Scattering}

To prove that the plaquette scattering in FeTe is magnetic in origin and determine its spin-space 
anisotropy to compare with the spin wave anisotropy measured in twinned samples \cite{Ysong18},
we carried out neutron polarization analysis at positions equivalent to ${\bf Q}_{\rm PL}=(\sfrac{1}{4},\sfrac{1}{4},0)$,
which sits directly on the $C_4$-symmetric ridge-like features and as far as possible from the
bicollinear Bragg positions ${\bf Q}_{\rm BC}=(\sfrac{1}{2},0,\sfrac{1}{2})$.
Our polarized neutron scattering experiments were performed on the IN20 triple-axis spectrometer at the Institut Laue-Langevin using largely the same setup as that of Refs. \cite{Ysong18,zhang14}.
3.5 grams of detwinned FeTe single crystals were mounted in the $[H,K,0]$ scattering plane with the $c$-axis vertical.
To remove background scattering, we eliminated the spring washers entirely from the aluminum clamp manifold and simply pressed firmly on the crystals using aluminum screws to press on the aluminum spacer plates holding the crystals in the manifold.
We find the detwinning ratio is at least 50\% (3:1 domain population) in this experiment, which we determine from energy scans at $Q_s=(\sfrac{1}{2},1,0)$ and $Q_w=(1,\sfrac{1}{2},0)$, which are the same as the bicollinear magnetic peak positions but projected into the $L=0$ scattering plane (and where bicollinear scattering intensity still appears due to the weak $L$ modulation).

\begin{figure}[t]
\includegraphics[scale=.5]{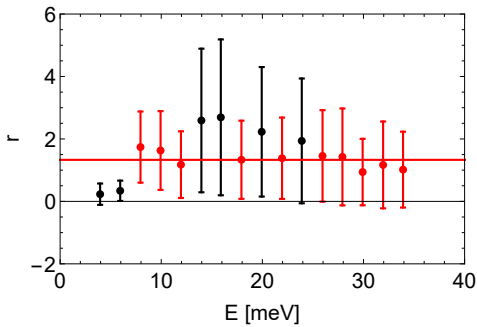}
\caption{
Intensity ratio between the points crystallographically equivalent to the plaquette position $Q_{PL}=(1/4,1/4,0)$.
}
\label{fig:rplot}
\end{figure}

\begin{figure*}[t]
\includegraphics[scale=.6]{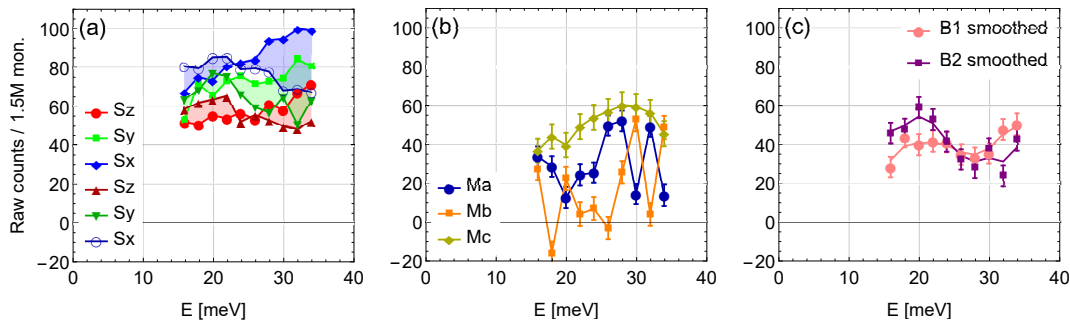}
\caption{
Background subtraction process for spin flip channel data at the plaquette positions ${\bf Q}_1=(3/4,3/4,0)$ [light colors in (a)] and ${\bf Q}_2=(1/4,7/4,0)$ [dark colors in (a)]. (b) and (c) show the calculated magnetic fluctuation projections $M_a$, $M_b$, and $M_c$ along the three crystallographic directions, and background values $B_1$ and $B_2$ at ${\bf Q}_1$ and ${\bf Q}_2$, respectively. In (c), the solid lines show the assumed actual background from smoothing the raw values.
}
\label{fig:pqbsplot}
\end{figure*}

\begin{figure}[t]
\includegraphics[scale=.5]{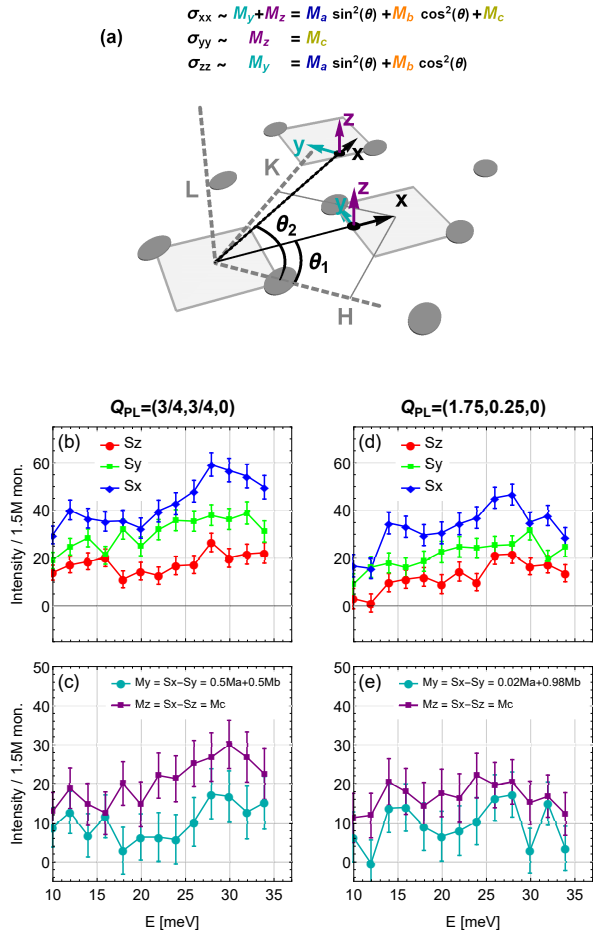}
\caption{
Smoothed-background-subtracted spin flip channel data at the plaquette positions ${\bf Q}_1=(3/4,3/4,0)$ and ${\bf Q}_2=(1/4,7/4,0)$, with positions and fluctuation directions defined in (a). The raw data in (b) and (d) are projected into the coordinate systems of the neutron spin polarization at each ${\bf Q}$ in (c) and (e), showing larger fluctuations along the $M_c$ direction, consistent with our expectations.
}
\label{fig:szpqout}
\end{figure}

Our analysis of the polarized neutron scattering data follows the method of Ref. \cite{zhang14}, described in the supplementary information therein.
Assuming that $M_a$, $M_b$, and $M_c$ are
spin excitations along the $a$, $b$, and $c$ axis directions, respectively [Figs. \ref{fig1}(a) and \ref{fig1}(d)], we can construct a matrix relating all scattering channel amplitudes to the magnetic fluctuation directions at two crystallographically equivalent $Q$.
This includes the assumption of an instrument-specific intensity ratio $r$ between the two equivalent ${\bf Q}$ positions as a function of energy, where we make the same assumption as in Fig. S1 in Ref. \cite{zhang14} that the value of $r$ is a constant after correcting for the magnetic form factor.
For the energy scans at plaquette positions, shown in Fig. \ref{fig:rplot}, with some values of $r$ considered outliers and excluded from the fit, we find $r=1.33$.
Using this value, we set up the full matrix for the values of $M_a$, $M_b$, $M_c$, 
and the energy-dependent background $B_1$ and $B_2$ at the two equivalent ${\bf Q}_{PL}$ positions,
and solve this overdetermined system with the least squares method.
Since the background is determined by the instrumental geometry and should change slowly with energy, we then filter the background values with a Gaussian kernel with a width of about 7 meV.
In Fig. \ref{fig:pqbsplot}, we show the results of this background determination: the left panel shows raw data scans in the $S_x$, $S_y$, and $S_z$ spin flip channels at the two ${\bf Q}$ positions; the center panel shows the (noisy) results of the moment determination; the right panel shows the determined background values with error and the smoothed fits.
After subtracting the smoothed background $B_1$ from all three spin flip channels at ${\bf Q}_1=(3/4,3/4,0)$, and $B_2$ from ${\bf Q}_2=(1/4,7/4,0)$, we can now calculate the amplitude of spin fluctuations in the coordinate system of the neutron spin polarization at each ${\bf Q}$, shown in Fig. \ref{fig:szpqout}(c) and (e), using the notation and color scheme of Ref. \cite{Ysong18} but with the scattering geometry of the present experiment.

Finally, we use the background-subtracted data to once again solve the full polarization matrix without the background terms by the method of least squares.
By projecting these moments along the crystallographic axes, this results in the final determination of $M_a$, $M_b$, and $M_c$, shown in 
Fig. \ref{fig1}(d).
Since the background is large and probably unreliable as high as $E=16$ meV, out of caution we only choose points with $E>16$ meV for this analysis except for the determination of $r$ as shown in Fig. \ref{fig:rplot}.

\begin{figure*}[t]
\includegraphics[scale=.45]{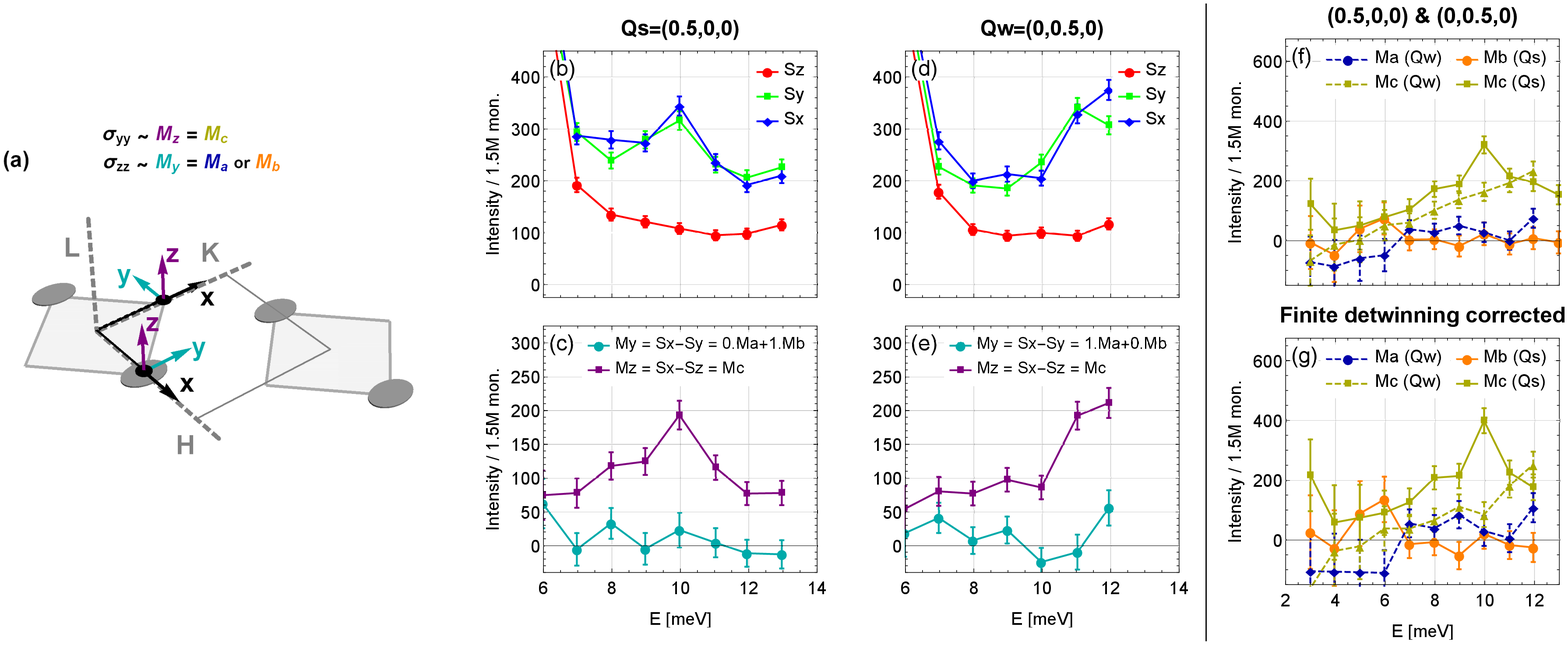}
\caption{
Magnetic moments at the BC position in the first Brillouin zone, with unique determination of some of the crystallographic moments due to the accidental alignment with the crystallographic axes. With the definition of the neutron spin coordinate systems in (a), the raw data in (b) and (d) at ${\bf Q}_1=(1/2,0,0)$ and ${\bf Q}_2=(0,1/2,0)$, respectively, are projected into the spin polarization coordinate systems in (c) and (e), revealing $c$-axis polarized fluctuations in both cases. This is further made evident by projecting the data into the crystallographic coordinate system in (f). After scaling the data based on the measured detwinning ratio, the projections show that fluctuations dominate at the majority-domain ${\bf Q}=(1/2,0,0)$, consistent with spin waves emanating from bicollinear order.
}
\label{fig:fzout}
\end{figure*}

By repeating this method for the bicollinear AF positions, we can extract the same determination of the moments in the polarization coordinate system.
In Fig. \ref{fig:fzout}, we demonstrate the determination of magnetic moments for the bicollinear positions in the first Brillouin zone, ${\bf Q}=(1/2,0,0)$, and $(0,1/2,0)$.
The projections from $M_y$ and $M_z$, in the coordinate system of the neutron spin polarization, to $M_a$, $M_b$, and $M_c$ along the crystallographic axes, is noted in the plot titles in Fig. \ref{fig:fzout}.
We correct this data for finite detwinning using the method of Ref. \cite{TChen}, wherein the finite detwinning ratio is scaled to reveal the fully detwinned behavior.
Our data shows predominantly $c$-axis fluctuations at the strong twin position, as expected.

\begin{figure}[t]
\includegraphics[scale=.5]{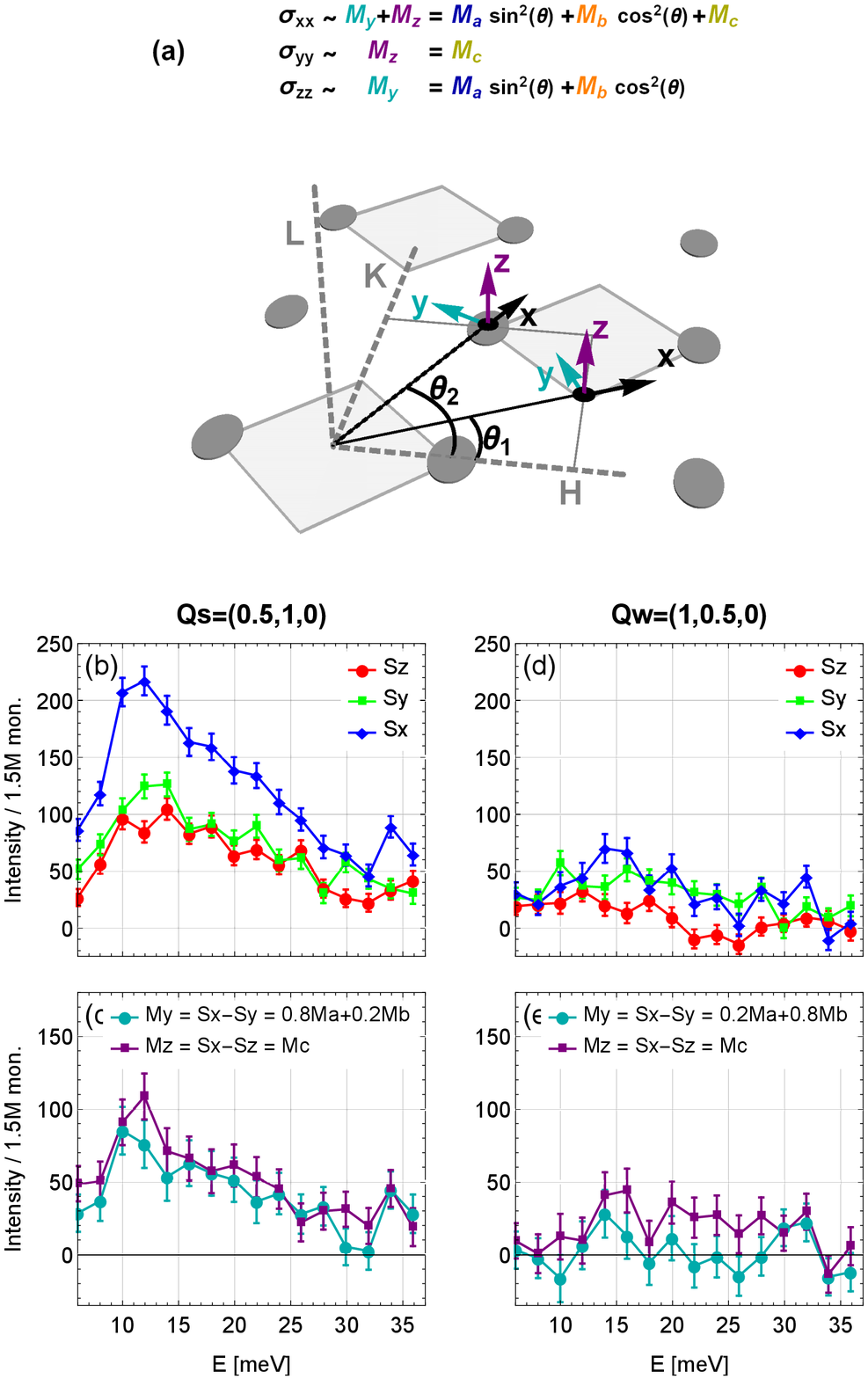}
\caption{
Magnetic fluctuations at the BC positions in the second Brillouin zone, from smoothed-background-subtracted spin flip channel data, analogous to Fig. \ref{fig:szpqout} at plaquette positions and extending the energy range of Fig. \ref{fig:fzout} at the bicollinear positions. Due to the different ${\bf Q}$ positions, the projections onto the neutron spin coordinate systems in (c) and (e), based on the smoothed raw data in (b) and (d), have a different projection onto the crystallographic axes $a$, $b$, and $c$ as noted in the insets to (c) and (e). In this figure, the data in all panels has been rescaled for the ideal case of 100\% detwinning, and therefore shows slightly $c$-axis polarized spin waves emanating from the majority-domain ``strong'' bicollinear order position ${\bf Q}_s=(1/2,1,0)$ only, and vanishing intensity above $E \sim$ 25 meV.
}
\label{fig:szbcout}
\end{figure}

In Fig. \ref{fig:szbcout}, we continue this analysis for bicollinear positions in the second Brillouin zone, ${\bf Q}=(1/2,1,0)$, and $(1,1/2,0)$.
Our data is consistent with Ref. \cite{Ysong18} in that it shows $c$-axis fluctuations.
After correcting for the finite detwinning, we can calculate $M_y$ and $M_z$, in the coordinate system of the neutron spin polarization.
The relationship to $M_a$, $M_b$, and $M_c$ is noted in the plot titles in Fig. \ref{fig:szbcout}, but determining the crystallographic projections is unreliable because we only measure one ${\bf Q}$ for each peak and have already used that information to correct the detwinning ratio.
The anisotropy of the bicollinear spin waves is not as strong as Fig. 3(a) in Ref. \cite{Ysong18}, because in our scattering plane the data is collected at $L=0$ and not at the $L=0.5$ or $L=1.5$ Bragg positions, and is therefore consistent with Fig. 3(b) in Ref. \cite{Ysong18}.

The energy dependence of spin excitation anisotropy at ${\bf Q}_{\rm PL}$ is similar to that 
the bicollinear ordering wave vector with $M_c>M_a>M_b$ [Fig. 1(d)].
The $C_4$ symmetric spin excitations above $E=26$ meV are essentially isotropic, consistent with the measurements on twinned samples \cite{Ysong18}.
Therefore, spin excitations of FeTe are anisotropic in both spin space and reciprocal (momentum) space below $E=26$ meV, and become isotropic at higher energies.

\begin{figure}[t]
\includegraphics[scale=.6]{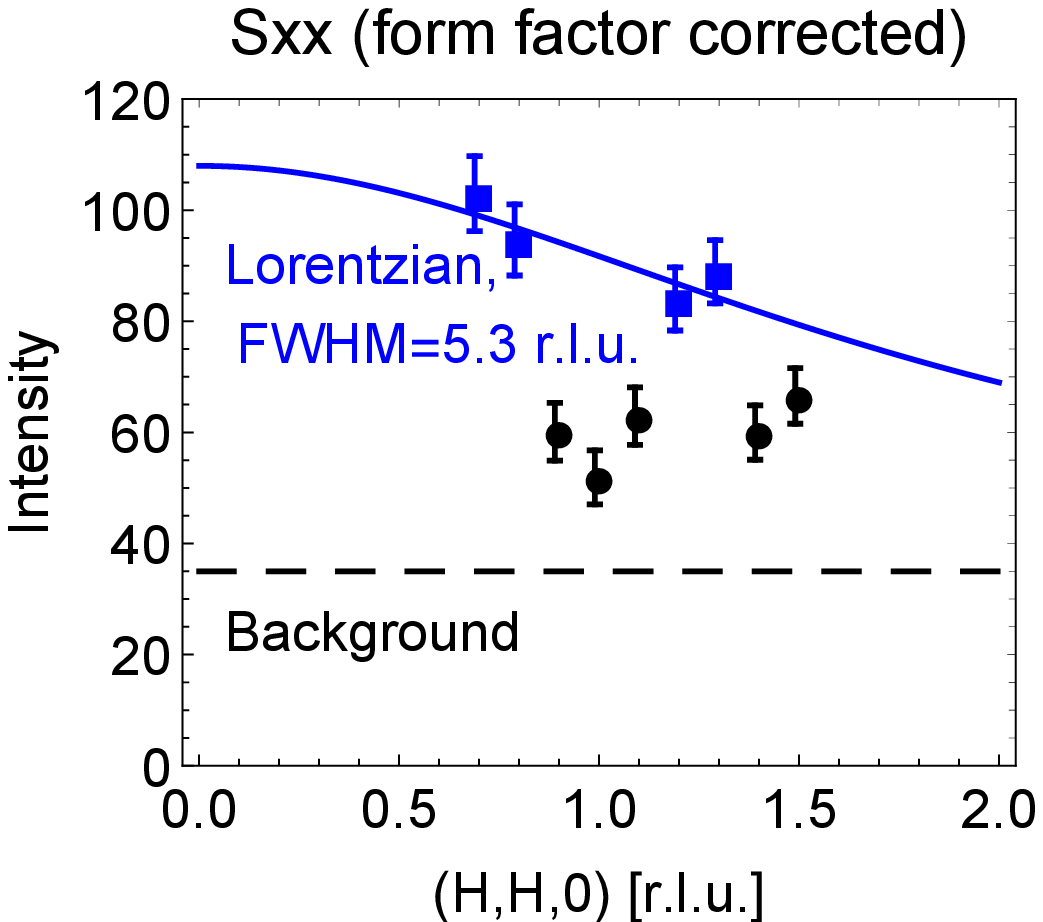}
\caption{
$S_{xx}$ channel spin-flip scattering in a longitudinal scan in the $[H,H,0]$ direction across the plaquette feature. Since this profile falls off faster than the magnetic form factor, it suggests a short-range nature of the plaquette excitations which is consistent with, but beyond the scope of, our long-ranged Hamiltonian model.
}
\label{fig:pqshortrangeplot}
\end{figure}

Finally, we find that spin waves at the plaquette positions are short-range in nature.
From longitudinal scans in the $[H,H,0]$ direction, we can compute the correlation length of the plaquette structures.
After correcting for the form factor of iron (Fe$^\text{2+}$), we fit a Lorentzian function centered at ${\bf Q}=0$, finding a FWHM of 3.7 r.l.u in the basal plane.
The scan points and fit for the $S_z$ spin flip channel is shown in Fig. \ref{fig:pqshortrangeplot}.

\section{Theory of bicollinear order and quasi-degenerate plaquette state in FeTe}

To understand the crossover at $E=26$ meV between these two behaviors, we consider a model of strongly frustrated local moments in which
a bicollinear ground state is nearly degenerate with a plaquette state \cite{Lai16}.
Our models are spin-1 $J-K$ ring-exchange models, 
where $J_n$ are the usual Heisenberg interactions $S_ i\cdot S_j$ and $K_n$ 
are the biquadratic interactions $(S_ i\cdot S_j)^2$, where the subscript $n$ 
means $n$th-neighbor coupling.

In this model, bicollinear and plaquette order are naturally degenerate and the degeneracy is broken in favor 
of the bicollinear state by a ring-exchange interaction and, due to the small size of this interaction,
it takes only small energy to produce spin excitations of the plaquette type even though the ground state is 
bicollinear \cite{Lai16}.

Following the theoretical work in Ref.~\cite{Lai16}, we consider the spin-$1$ bilinear-biquadratic model whose Hamiltonian is defined as
\begin{eqnarray}
 H&=&  \sum_{i, j} \left[ J_{ij} {\bf S}_i \cdot {\bf S}_j + K_{ij} \left( {\bf S}_i \cdot {\bf S}_j \right)^2 \right] \nonumber,
\label{ham_ring}~~~~~
\end{eqnarray}
where ${\bf S}_{i}$ is the local spin-$1$ moment at site $i$ of the Fe square lattice, and the bilinear and biquadratic interactions are chosen up to the third neighbors. 

Within a large parameter regime, the ground state bicollinear order discovered in the experiments and the plaquette state  with the same ordering wave vector are quasi-degenerate, and it is suggested that the presence of ring exchange terms can lift the degeneracy between  bicollinear and plaquette states. At finite energy, it is expected that both should contribute to the dynamical spin structure factor $S({\bf Q},E)$ probed by the neutron scattering measurements. 

For obtaining the $S({\bf Q}, E)$ of bicollinear and plaquette states, we utilize the linear spin wave theory to re-express spin-$1$ operator in terms of Holstein-Primakoff bosons as
\begin{eqnarray}
 S^+_j = \sqrt{2M - b_j^\dagger b_j}~b_j,&~S^-_j = (S^+_j)^\dagger,&~ S^z_j= M - b_j^\dagger b_j, ~~~
\end{eqnarray}
where $M$ is the value of the spin, and here $M= 1$. For sites $j$ with $|S^z = +1\rangle$, we can simply rewrite spin operator $S_j$ in terms of bosons directly. For sites $j$ with $|S^z = -1\rangle$, we first perform a $\pi$ rotation around $S^x$, and, therefore, $S_j^\pm \rightarrow S_j^\mp$, and $S_j^z \rightarrow - S_j^z$. After the rotation, we re-write the spin operator in terms of bosons. 

Within the linear spin wave theory, we then proceed to calculate $S({\bf Q}, E)$ for both bicollinear and plaquette states. The dynamical spin structure factor can be extracted based on the formula:
\begin{widetext}
\begin{eqnarray}
S^{\mu \nu}({\bf Q},E) &=& \frac{1}{N_s} \sum_{j,\ell} e^{i \bm{q}\cdot (\bm{r}_j  - \bm{r}_\ell )} \sum_f \left\langle S^\mu_j | f \right\rangle \left\langle f | S^\nu_\ell \right\rangle \delta \left(\omega - E_f+ E_G\right) \\
&=& \frac{1}{ n N_u} \sum_{\alpha, \beta} \sum_f \bigg{\langle} \sum_{j \in \alpha} S^\mu_j e^{i \bm{q} \cdot \bm{r}_j} \bigg{|} f \bigg{\rangle}  \bigg{\langle} f \bigg{|} \sum_{\ell \in \beta} e^{-i \bm{q} \cdot \bm{r}_\ell}S^\nu_\ell \bigg{\rangle} \delta \left( \omega - E_f + E_G \right) \\
&=& \frac{1}{n} \sum_{\alpha, \beta} \sum_f \left\langle S^\mu_\alpha (\bm{q}) | f \right\rangle \left\langle f | S^\nu_\beta (- \bm{q}) \right\rangle \delta \left( \omega - E_f + E_G \right),
\end{eqnarray}
\end{widetext}
where we abbreviate $| S^\nu_\ell \rangle \equiv S^\nu_\ell |G\rangle$, where $|G\rangle$ represents the ground state,  $|f\rangle$ represents the intermediate state,  $\mu$, $j,~\ell$ represent the site labelings, $N_s,~N_u$ represent the number of sites and number of unit cells, $\mu, \nu = x,~y~z$ represent the component of the spin operator, $n = 4~(16)$ for BC (PL) order, and $\alpha, \beta$ represent the sublattice labelings within each unit cell. For accommodating the energy gap between the ground state bicollinear and the quasi-degenearate plaquette states, we shift the origin of the $S^{\mu\nu}({\bf Q}, E)$ of plaquette state to a small finite energy $E_0$ and combine the $S^{\mu\nu}({\bf Q},E)$ of bicollinear and plaquette states at equal weight ($50\%$ BC +$ 50\%$ PL). For the isotropic system, we find $S^{xx}({\bf Q}, E) = S^{yy}({\bf Q},E),~S^{zz}({\bf Q}, E) = 0$, and obtain $S({\bf Q}, E) = \sum_{\alpha = x, y} S^{\alpha \alpha}({\bf Q}, E)$.

\begin{figure}[t]
\includegraphics[scale=.42]{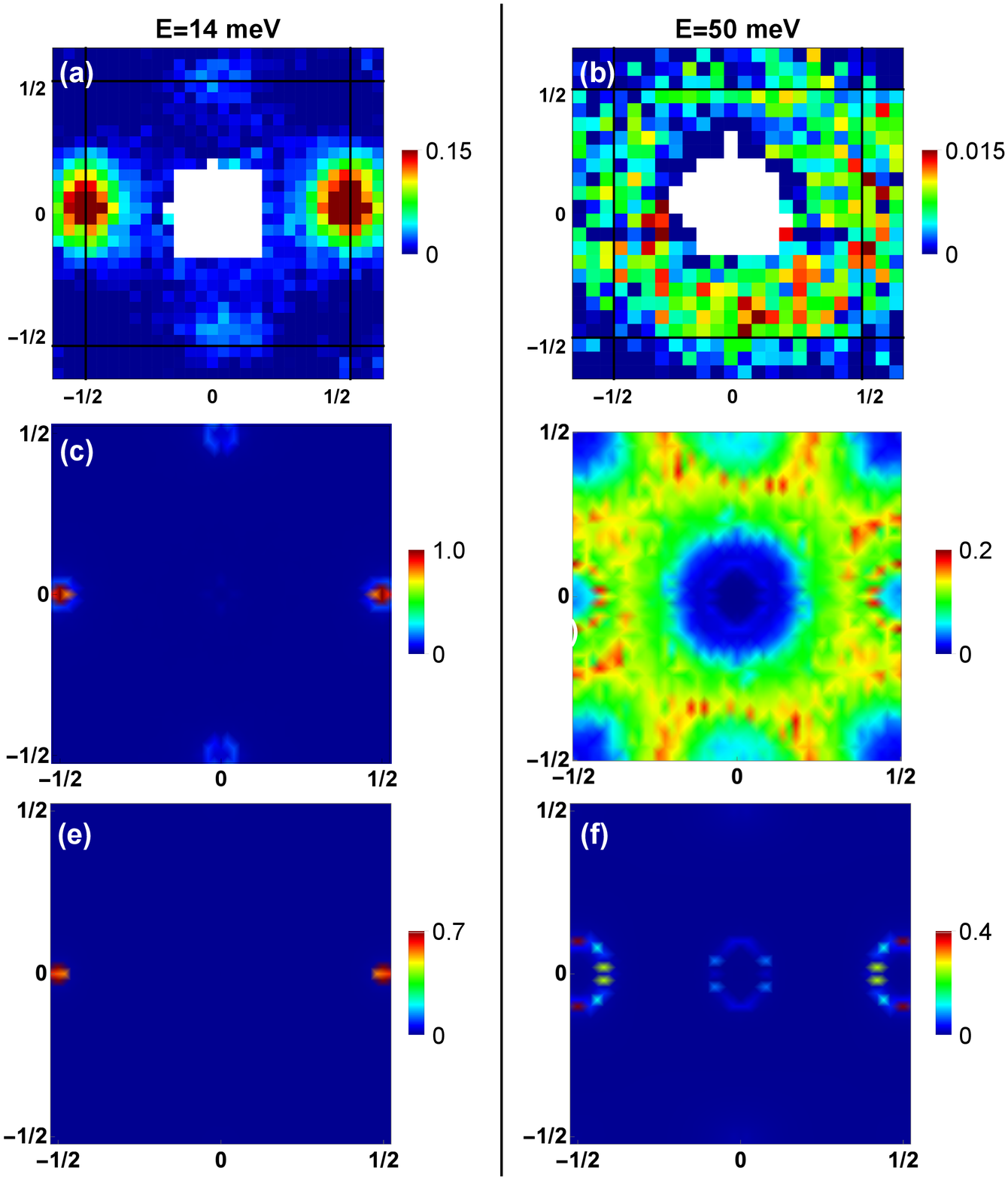}
\caption{
(a-b) Neutron scattering data at $E=14$ ($E_i=30$) and $E=50$ meV ($E_i=80$), with the central magnetic zone from Fig. \ref{fig:tof-lowE}(b) shown in black.
(c-d) $S({\bf Q},E)$ calculated with the bicollinear/plaquette biquadratic Hamiltonian, at (c) $\omega$=1.7 (corresponding to approximately 20 meV) and (d) $\omega$=7.5 (approximately 90 meV), showing the correspondence with (a-b). The correspondence between $\omega$, the energy parameter in the calculations, and experimental energy is made by identifying $\omega$=20 as the magnetic band top which lies near 250 meV \cite{LCFP11}.
(e-f) $S({\bf Q},E)$ calculated with the bicollinear biquadratic Hamiltonian only, at (e) $\omega$=1.7 and (f) $\omega$=7.5, showing a disagreement with (a-b).
}
\label{fig:exp-th}
\end{figure}

For illustration, we take the parameters $(J_1, J_2, J_3, K_1, K_2, K_3) = (1, 0.8, 1, -0.8, 0.1, -0.8)$ within the parameter regime, where bicollinear and plaquette state are both stable and quasi-degenerate as illustrated in Ref.~\cite{Lai16}, to obtain the dynamical spin structure factor $S({\bf Q}, E)$ shown in Fig. \ref{fig:exp-th}(c-d).
The delta function is replaced by a Lorentizian function with a finite width of width $\delta\omega=0.2$ (approximately 2.5 meV) to capture the damping produced by itinerant electrons.
These images capture the salient features of the neutron scattering data shown in Fig. \ref{fig:exp-th}(a-b) at low ($E=14$ meV) and high energy ($E=50$ meV).
In Fig. \ref{fig:exp-th}(e-f), we show the patterns resulting from a purely bicollinear ground state only
(generated by small changes in the method of solution \cite{Lai16})
, resulting in twofold-symmetric spin waves generated from the double-stripe bicollinear order alone that do not agree with the observed neutron data.
The good agreement with the bicollinear/plaquette Hamiltonian 
suggests that the bicollinear and plaquette states in this material are competing at a range of energy scales 
throughout the magnetic bandwidth, consistent with true magnetic frustration.

\section{Conclusions}

By comparing our neutron scattering results with calculations in the frustrated spin model,
 we conclude that FeTe has nearly degenerate bicollinear and plaquette order 
arising from localized spins. 
Our results are consistent with polarized neutron scattering experiments
on twinned \cite{Ysong18} and detwinned FeTe.
Together, these experiments would suggest that the measured spin excitation spectrum originates from
localized moments on the Fe sites, corresponding to a combination of isotropic paramagnetic scattering 
with $C_4$ symmetry from the plaquette correlations and the transverse spin waves of the bicollinear order without phase separation.

Iron chalcogenides other than FeTe can have multiple ground states with similar energies 
and suggest local moment physics is the dominant paradigm throughout this family of compounds.
In superconducting FeSe ($T_c\approx 8$ K) \cite{Hsu08}, 
which is isostructural to FeTe and has similar lattice distortion but without static AF order, 
spin excitations are found to be frustrated by competing collinear and N${\rm \acute{e}}$el magnetic instabilities, 
which occur at wavevectors $(\pm \sfrac{1}{2},\pm \sfrac{1}{2})$ and $(\pm 1,0)/(0,\pm 1)$ in this notation, 
respectively \cite{WSPZ16,Fwang15,Glasbrenner,RYu15}. Superconductivity in detwinned FeSe induces a neutron spin resonance that has $C_2$ rotational symmetry, suggesting 
orbital selective electron pairing \cite{TChen}. 
On the other hand, 
 the lower-$T_c$ ($=4$ K) isostructural analog FeS
 has strictly collinear magnetic correlations at finite energy with no evidence for superconductivity-induced neutron spin resonance \cite{LZWW15,KLTB16,HPGK16,MGZS17}.
 Finally, we note again that our model is relevant to the macroscopic FeTe layers and does not depend on the existence of interstitial Fe sites, although it will be important to verify that these observations can be verified with future experiments on FeTe with controlled interstitial disorder such as has been achieved in FeTe$_{1-x}$Se$_{x}$ with the recently developed technique of Te-annealing \cite{XSYZ18,YSPT16,LGGW15,OHKA19}.
Therefore, magnetic frustration from localized moments plays a significant role 
in determining the electronic properties of FeTe and its closely related superconducting compounds, 
and is of key importance to understand the ground state of iron chalcogenides.

\section{acknowledgments}

The neutron scattering work at Rice is supported by the U.S. DOE BES under contract no. DE-SC0012311 (P.D.). 
The materials synthesis efforts at Rice and PSU are supported by the Robert A. Welch Foundation Grant Nos. C-1839 (P.D.) and the U.S. DOE under grant DE-SC0019068 (Z.Q.M.), respectively.   
The research at ORNL was sponsored by the Scientific User Facilities Division, Office of BES, U.S. DOE.
The theory work at Rice is supported by  the 
U.S. DOE, BES, under Award No. DE-SC0018197
 (H.-H.L. and Q.S.), Robert A.Welch Foundation Grant No. C-1411
and a Smalley Postdoctoral Fellowship of the Rice Center for Quantum Materials (H.-H.L.).

\end{document}